\documentclass[aps,prl,showpacs,showkeys,twocolumn,groupedaddress,superscriptaddress]{revtex4}
\usepackage{graphicx}%
\usepackage{dcolumn}% Align table columns on decimal

\begin{document}

\title{Formation  of a Direct Kolmogorov-like Cascade
\\of Second Sound Waves in He~II}

\author{G.~V.~Kolmakov}%
%\email{g.kolmakov@lancaster.ac.uk}
\affiliation{Department of Physics, Lancaster University,
Lancaster, LA1 4YB, UK}\affiliation{Institute of Solid State
Physics RAS, Chernogolovka, Moscow region, 142432, Russia}

\author{V.~B.~Efimov}
\affiliation{Department of Physics, Lancaster University,
Lancaster, LA1 4YB, UK}\affiliation{Institute of Solid State
Physics RAS, Chernogolovka, Moscow region, 142432, Russia}

\author{A.~N.~Ganshin}
\affiliation{Department of Physics, Lancaster University, Lancaster,
LA1 4YB, UK}

\author{P.~V.~E.~McClintock}
\affiliation{Department of Physics, Lancaster University, Lancaster,
LA1 4YB, UK}

\author{L.~P.~Mezhov-Deglin}
\affiliation{Institute of Solid State Physics RAS, Chernogolovka,
Moscow region, 142432, Russia}

\date{
%\textbf{Accepted for publication in Phys. Rev. Lett.}
\today
}% It is always \today, today,
       % but any date may be explicitly specified

\begin{abstract}

Based on measurements of nonlinear second sound resonances in a
high-quality resonator, we have observed   a steady-state wave
energy cascade in He II involving a flux of energy through the
spectral range towards high frequencies. We show that the energy
balance in the wave system is nonlocal in $K$-space  and that the
frequency scales of energy pumping and dissipation are widely
separated. The wave amplitude distribution follows a power law
over a wide range of frequencies. Numerical computations yield
results in agreement with the experimental observations. We
suggest that second sound cascades of this kind may be useful for
model studies of acoustic turbulence.

\end{abstract}



\maketitle

%{\it Introduction.}
Turbulence in the system of waves in a nonlinear non-dispersive
medium is usually referred to as acoustic turbulence, or Burgers
turbulence (BT) in honor of the Burgers equation which provides a
useful description of the phenomenon. BT has been at the focus of
numerous investigations during the last few decades because of its
importance for basic nonlinear physics and in view of numerous
applications in engineering and fundamental science
\cite{Zakharov,falk1,falk2}. Well known examples of BT include the
turbulence of sound waves in oceanic waveguides \cite{at1},
magnetic turbulence in interstellar gases \cite{at2}, and  shock
waves in the solar wind and their coupling with the Earth's
magnetosphere \cite{at3}.

In this Letter we propose  a novel approach to the experimental
study of BT: the use of nonlinear second sound standing waves in
superfluid $^4$He (He~II) within a high-quality resonator. Second
sound is a slightly dissipative temperature/entropy wave that
arises in superfluids and perfect crystals \cite{Landau}. Second
sound in He~II is characterized by nonlinear properties that are
rather strong compared to those of ordinary sound (pressure/density
waves) in conventional solids, liquids and gases
\cite{Fairbank,knz,Ahlers,Borisenko}. The velocity $u_2$ of second
sound depends on its amplitude and, to a first approximation, can
be written as
\begin{equation}
u_2 = u_{20} (1 + \alpha \delta T),
\end{equation}
where $u_{20}$ is the wave velocity at negligibly small  amplitude,
$\delta T$ is the wave amplitude, and
\[
\alpha = {\partial \over \partial T} \ln \left(
u_{20}^2 {C \over T} \right),
\]
is the nonlinearity coefficient of second sound. $C$ is the heat capacity
per unit mass of liquid helium at constant pressure, and $T$ is the
temperature.

There are huge advantages in the use of roton second sound
waves in He~II for model studies of nonlinear wave interactions.
Within the experimentally convenient temperature range 2.17--1.5 K
the nonlinearity coefficient $\alpha$ can be tuned, just by
changing the bath temperature:  $\alpha \rightarrow -\infty$ near
the normal-to-superfluid transition at $T_{\lambda} =2.17$  K;
$\alpha$ passes through zero  at $T_{\alpha}=1.88$ K; and $\alpha
\sim +2$ K$^{-1}$ at $T \sim 1.5$ K  \cite{Fairbank}. Thus one can
study the dynamics of both nearly-linear and  strongly-nonlinear
waves with positive (like conventional sound) or  negative
nonlinearity while using exactly the same experimental techniques.
Such possibilities are unavailable in conventional experiments.
The fact that the velocity of second sound $u_{20} \leq 20$ m/s is
more than an order of magnitude less than the velocity of
conventional sound in gases and in condensed media allows us to
increase the time resolution of the measurements.

We use a thin-film heater to generate the second sound, and a
thin-film thermometer (a low-thermal-inertial superconducting
bolometer) as a detector. Use of a high-$Q$ resonator enables us
to create nonlinear standing second sound waves of high amplitude
with only small heat production at the source. The wave amplitude
$\delta T$ can be changed from 0.05 mK up to a few mK, so that the
Mach number can reach ${\rm M} = \alpha \delta T \sim 10^{-4} -
10^{-2}$, and the proper Reynolds number, defined \cite{s1}  as
${\rm Re} = \alpha u_{20} (\partial \delta T /
\partial x)  / \gamma_{\omega}   \sim \alpha Q \delta T $, can
be changed from 1 to 90 (here $\gamma_{\omega}$ is the damping
coefficient of a second sound wave of frequency $\omega$,
evaluated from the $Q$-factor of the resonator). It allows us to
study, for the first time, the transition from the linear regime
to turbulence in the system of second sound waves as the driving
amplitude is increased. The experimental arrangements were similar
to those used earlier \cite{Borisenko}. The resonator was formed
by a cylindrical quartz tube of length $L=7$ cm and inner diameter
$D=1.5$ cm. The film heater and bolometer were deposited on the
surfaces of  flat glass plates capping the ends of the tube. The
heater was driven by an external sinusoidal voltage generator in
the frequency range between 0.1 and  100 kHz: The frequency of the
second sound (at twice the frequency of the voltage generator) was
set equal to one of the longitudinal resonant frequencies. The
measurements reported below were conducted at $T=2.079$ K where
$\alpha \approx -7.6$ K$^{-1}$, and at $T=1.775$ K where $\alpha
\approx 1.8$ K$^{-1}$. The $Q$-factor of the resonator determined
from the widths of longitudinal resonances at small heat flux
densities was $Q \sim 1000$ for resonant numbers $p \leq 10$ and
$Q \sim 3000$ for $30<p<100$. The steady-state second sound
distribution in the resonator was Fourier-analysed and its power
spectrum was computed.

\begin{figure}[t]
\centerline{\includegraphics[width=70.mm]{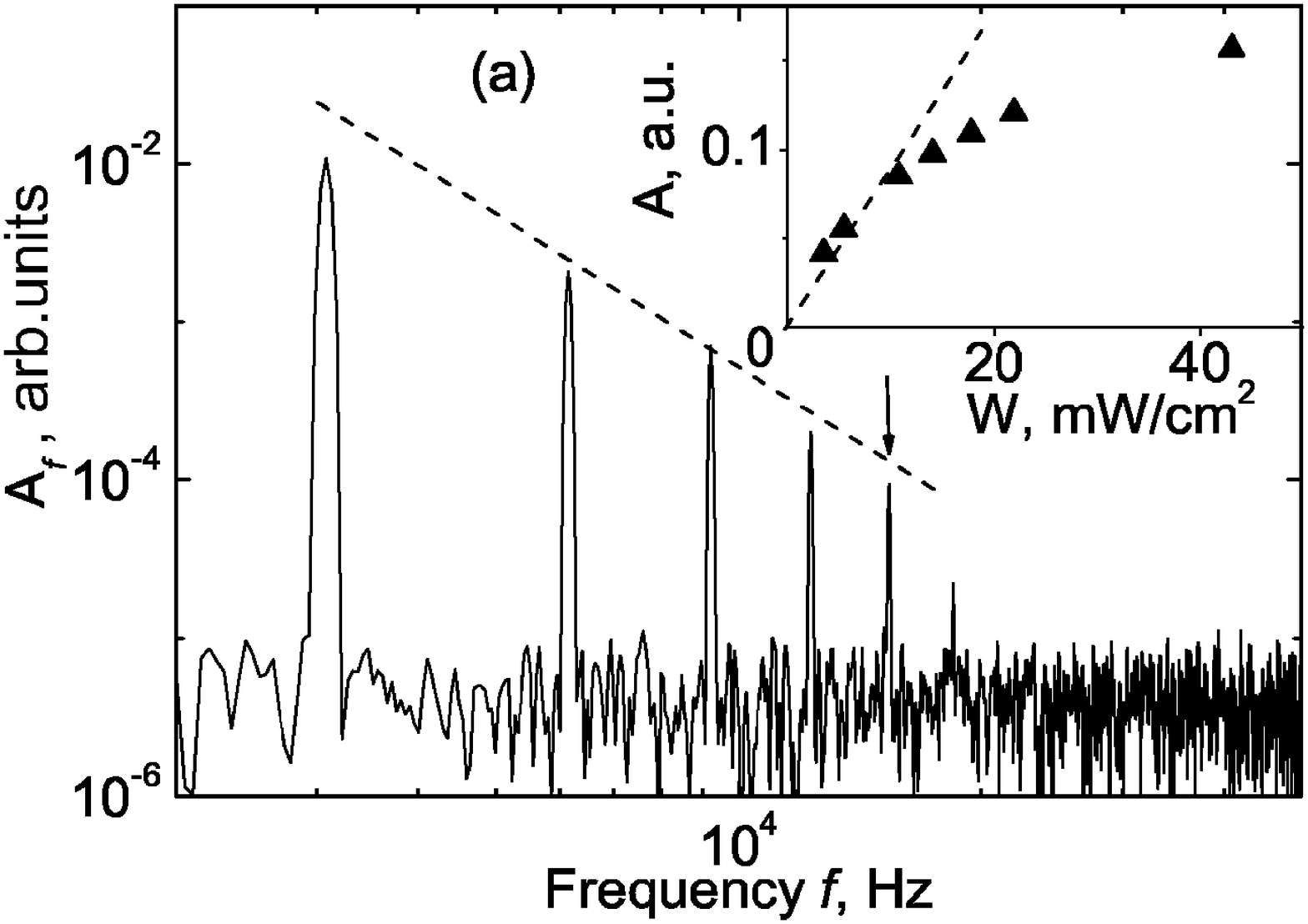}}
\centerline{\includegraphics[width=70.mm]{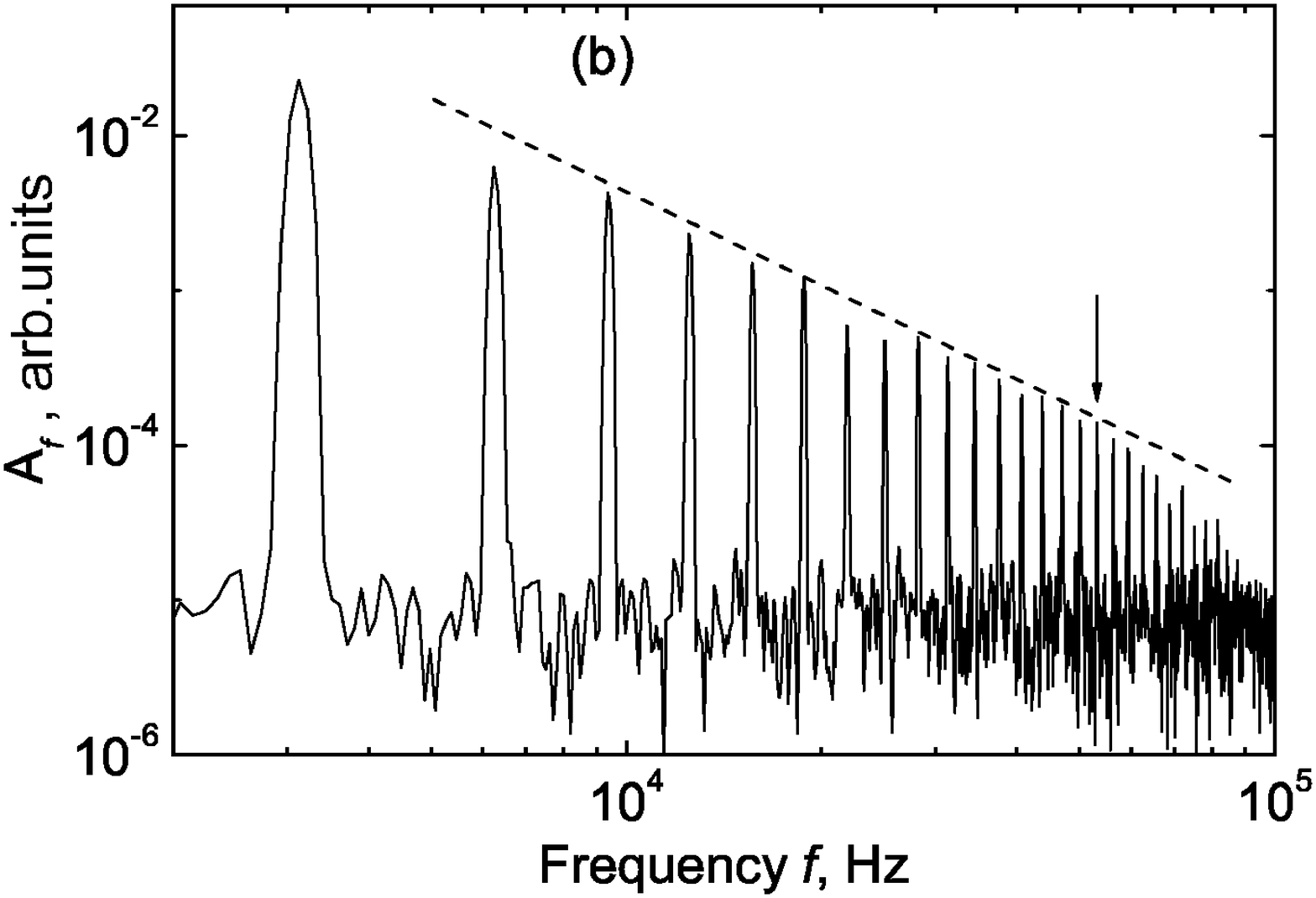}}
\caption{Power spectral amplitudes $A_f$ of standing waves
recorded at $T=2.079$ K when driving at the 31st resonant
frequency, $f_d=3130$ Hz. The ac heat flux density from the heater
was (a) $W=5.5$ and (b) 22 mW/cm$^2$. The dashed line in (a) is a guide
to the eye; that in (b) corresponds to $A_f \propto f^{-1.5}$. The
arrows indicate  positions of the viscous cutoff frequency $f_b$  in
each case. In the inset triangles show the observed dependence of the
wave amplitude $A$ on ac heat flux density $W$, measured for 31st
resonance, and the dashed line corresponds to the linear
dependence $A \propto W$ valid at small $W$. } \label{fig1}
\end{figure}

Fig.\ 1 shows some typical experimental results at 2.079~K. For a
small ac heat flux density $W<1$ mW/cm$^2$, corresponding to a
standing wave amplitude less than 0.5 mK and to Reynolds numbers
${\rm Re} \sim 1$, we observed a nearly linear regime of wave
generation: the amplitude $A$ of the  standing wave was
proportional to the heat flux density $W$ (see Fig.\ 1 inset).
Increase of the excitation above a few mW/cm$^2$  (${\rm Re} >
10$) led to large deviations from $A \propto W$ and to visible
deformation of the initially sinusoidal standing wave, accounted
for by the formation of multiple harmonics in its spectrum, as
shown in Fig.\ 1 and the inset. At $W \sim 80$ mW/cm$^2$  we
observed a marked reduction of the $Q$-factor to $\sim 500$ for
longitudinal resonances with $p \sim 30$. This may be associated
with e.g.\  broadening of the resonance curve due to  vortex
creation in the bulk He~II. In what follows, we discuss only the
measurements made at $W < 80$ mW/cm$^2$.

It is clearly evident from Fig.\ \ref{fig1} that the main spectral
peak lies at the driving frequency $f_d$, and that high-frequency
peaks appear at its harmonics, $f_n=f_d \times n$ with
$n=2,3,\ldots$ It can be seen that a cascade of waves is formed
over the frequency range up to 15 kHz in Fig. \ref{fig1}(a), and
up to 80 kHz in Fig. \ref{fig1}(b), i.e.\ up to frequencies 5 and
25 times higher than the driving  frequency. As shown in Fig.
\ref{fig1}(b), the dependence of peak height on frequency may be
described by a power-like law  $A_f = {\rm const} \times f_n
^{-s}$ at frequencies lower than  some cutoff  frequency $f_b$.
For sufficiently high heat flux densties $W>10$ mW/cm$^2$ the
scaling index tends to $s\approx 1.5 \pm 0.3$.

\begin{figure}[t]
\vspace{3.mm}
\centerline{\includegraphics[width=70.mm]{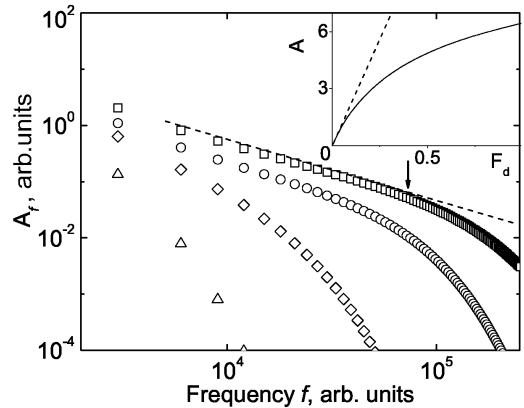}}
%\centerline{\includegraphics[width=80.mm]{Fig1b.eps}}
\caption{Second sound power spectral amplitudes $A_f$ calculated
numerically from Eq.\ (2) for four different driving force
amplitudes: $F_d$ = 0.01 (triangles); 0.05 (diamonds); 0.1
(circles); and 0.3 (squares). The dashed line corresponds to
$A_{f} \propto f^{-1}$ as predicted by the theory
\cite{falk1,Kuzia}. The arrow marks the boundary frequency $f_b$
of the inertial range for $F_d = 0.3$. The inset shows the
calculated dependence on $F_d$ of the standing wave amplitude $A$:
solid line -- nonlinear waves, $\alpha <0$ ($T>T_{\alpha}$);
dashed line -- linear waves, $\alpha =0$ ($T=T_{\alpha}$). }
\label{fig3}
\end{figure}

In order to understand how the wave cascade is formed, we have
undertaken a numerical study of the dynamics of nonlinear second
sound waves within the high-$Q$ resonator. Full details will be
given elsewhere, but in essence it involves direct integration of
the two-fluid thermohydrodynamical equations \cite{Landau},
expanded up to quadratic terms in the wave amplitude. We use a
representation of the second sound waves in terms of Hamiltonian
variables \cite{Pokrovsky,Kolmakov}, permitting us to improve the
precision of the calculations. The numerical technique is  similar
to that used in earlier  studies \cite{falk1}  of developed
two-dimensional acoustic turbulence. In the present model,
however, wave damping was taken into account at all frequencies, a
feature that is of key importance \cite{prl} for a correct
description of the formation of a cascade of nonlinear sound waves
with increasing driving force amplitude. The numerical results
obtained are suitable for direct comparison with our experimental
observations shown in Fig.~1.

The equations of motion governing non-local
energy balance in the system read
\begin{eqnarray}
i {\partial {b}_n \over \partial t} = \sum_{n_1, n_2} V_{n,n_1,n_2}
(b_{n_1}
b_{n_2}
\delta_{n-n_1 -n_2} \nonumber \\
 + 2 b_{n_1} b_{n_2} ^* \delta _{n_1 - n_2 -n} )
- i \gamma_n b_n + F_{d},
\end{eqnarray}
where $b_n (t) = (1/2) ( B_n^{-1} S_n + i B_n \beta_n)$ is the
time-dependent canonical amplitude of second sound at the $n$-th
resonant frequency $f_n$; $S_n$ and $\beta_n$ are the space
Fourier components $\sim \cos(2 \pi f_n x/u_{20} L)$ of the
entropy and of the potential of the normal-superfluid relative
velocity; $B_n=(f_n C/T)^{1/2}$; $V _{n,n_1,n_2} \propto \alpha (n
n_1 n_2)^{1/2}$ describes the three-wave interaction; $\gamma_n =
\nu n^2$ models the viscous damping of second sound; $F_{d}
\propto W $ is the amplitude of the force driving the $n$-th
resonant mode. We neglect the coupling between temperature and
pressure oscillations in view of the smallness of the thermal
expansion of He~II in the relevant temperature range. The wave
spectrum was calculated as $A_{f} \propto B_n \, (b_n + b_n^*)$.
We used a periodic driving force of frequency equal to a resonant
frequency of the resonator corresponding to the conditions of the
measurements, taken as $f_d=3000$ in dimensionless units. The
effective viscosity coefficient $\nu$ was fitted to the measured
value of the quality factor $Q \sim 3 \times 10^3$ of our
resonator, to facilitate comparison of the model results with
those from the experiments.

Fig.\ \ref{fig3} shows the evolution of the steady-state spectrum
with increasing driving force amplitude $F_{d}$, calculated for
$\alpha < 0$ ($T>T_{\alpha}$). Points on the plot correspond to
the amplitudes of the peaks in the spectrum. It is seen that, at
small driving amplitude  $F_d \sim 0.01$ (triangles), viscous
damping prevails at all frequencies  and  a turbulent  cascade is
not formed: the amplitude of the second harmonic is an order less
that the amplitude of the main harmonic. In this regime the wave
shape is close to linear. At intermediate driving amplitude $F_d
\sim 0.05$  (diamonds) nonlinearity starts to play a role at
frequencies of the order of driving frequency, and a few harmonics
are generated (cf.\ Fig.\ 1(a)). At high driving amplitudes $F_d
\geq 0.1$ (circles and squares) a well-developed cascade of second
sound waves is formed up to frequencies 30 times higher than the
driving frequency, i.e.\ the behaviour predicted by the
computations is very similar to that observed in our experiments
(Fig.\ 1(b)).

Formation of the spectra observed in the experiment and numerics
is evidently attributable to the cascade transfer of wave energy
through the frequency scales due to nonlinearity, thus
establishing an energy flux in $K$-space directed from the driving
frequency towards the high frequency domain. Following the basic
ideas formulated in Refs.
\cite{Kolmogorov,Zakharov1,Zakharov,Kuzia}, such a highly excited
state of a system with numerous degrees of freedom is referred to
as turbulent. So we may infer that at relatively high driving
amplitudes we  observe  {\it acoustic turbulence} formed in the
system of second sound waves within the inertial (non-dissipative)
range of frequencies. Formation of the observed direct cascade is
qualitatively  similar to creation of the Kolmogorov distribution
of fluid velocities over frequencies in the bulk of classical
incompressible fluids \cite{Landau}, or in a system of surface
waves \cite{Diachenko,prl}. But we found both in experiment and
numerics that the amplitudes of high frequency harmonics rise
gradually  with the increase  of the driving amplitude.

In some range of frequencies the calculated spectrum obeys a power
law with $s \approx 1$ (squares in Fig.\ 2) that is in qualitative
agreement with the theory \cite{falk1,Kuzia}. The difference in
slope of the calculated spectrum from the value $s = 1.5 \pm 0.3$
observed in the experiment may be attributable to our resonator
having an insufficiently high quality factor, as well as to the
influence of other mechanisms of wave energy relaxation not taken
into account in our computations. The precise identification of
the scaling index characterising a turbulent distribution is
obviously important for connecting BT theory to real physical
experiments. To clarify the origin of the difference between
observations and computations, we plan to make further
measurements and to undertake a systematic study in the vicinity
of $T_{\lambda}$, where the nonlinearity of second sound rises
much faster than its damping \cite{Ahlers}, and the proper Mach
and Reynolds numbers should therefore be substantially increased.

From a comparison of the measurements and computations (Figs. \ref
{fig1} and  \ref {fig3},  insets) we can conclude also that the
deviations from the linear dependence $A\propto W$ observed at
moderate driving amplitudes is caused by nonlinear outflow of the
energy to higher frequency scales, eventually reaching the
dissipative frequency range, rather than by the scattering of
second sound from quantized vortices in He~II, as considered
earlier \cite{Ahlers,Nemirovskii}.

The effect of the high frequency cutoff  of the wave spectrum
manifests itself as an abrupt decrease in the amplitudes of the
harmonics at $W< 10$ mW/cm$^2$ (compare Fig.~\ref{fig1}(a) and
Fig.~\ref{fig3} (triangles)), and as a change of slope on
double-log scales, at higher  $W$ (cf.\ Fig.~\ref{fig1}(b) and
Fig.~\ref{fig3} (circles and squares)), at some frequency $f=f_b$.
At $f \sim f_b$ the nonlinear mechanism for nearly non-dissipative
transfer of the wave energy changes to viscous damping of the
waves  (cf.\ observations of the high-frequency edge of the
inertial range of frequencies  of capillary turbulence on the
surface of liquid hydrogen \cite{prl}). It causes a faster
reduction of sound amplitude at frequencies $f>f_b$, as observed.

\begin{figure}[t]
\centerline{\includegraphics[width=70.mm]{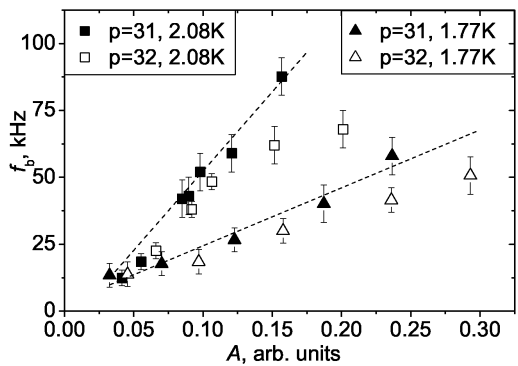}}
%\centerline{\includegraphics[width=80.mm]{Fig1b.eps}}
\caption{Dependence of the viscous cutoff frequency $f_b$ on the
amplitude $A$ of the standing wave for different temperatures,  for
resonance numbers $p=31$ and 32.
Dashed lines: the results of our  computations based on
the theory \cite{Kuzia}; data points:  experimental measurements.}
\label{fig2}
\end{figure}

The dependence of the boundary frequency $f_b$ on the standing
wave amplitude $A$ is shown in Fig \ref {fig2}. It is clearly
evident that the inertial range is extended towards higher
frequencies when the driving amplitude is increased. At
sufficiently large amplitudes $A$, the boundary frequency $f_b =
{\rm const} (T,f_d) \times A$, which agrees with our calculations
based on \cite{Kuzia}. Such a dependence was observed when driving
at resonant frequencies with odd resonant numbers $p$ (the full
symbols in Fig. \ref{fig2}).

When driving at even $p$ (open symbols), however, the boundary
frequency is noticeably lower than that measured for the nearest
odd resonant number, for  $W>10$ mW/cm$^2$. This reduction may be
connected with  a change in the mechanism of energy relaxation in
the wave system caused by the generation of subharmonics with
frequencies lower than $f_d$. We recently detected the appearance
of subharmonics at high $W$ when driving at even $p$, a phenomenon
that promises to be of huge  interest for nonlinear sound wave
dynamics \cite{s1,s2}. We plan to study it in detail in the near
future.

In conclusion, we have demonstrated that the system of roton
second sound waves in a high-$Q$ resonator filled with He~II can
be used as an effective tool for the detailed modelling and
investigation of acoustic turbulence. We observed for the first
time a smooth crossover  in the system of second sound waves from
nearly-linear regime at low driving amplitudes (at ${\rm Re} \sim
1$) to a nonlinear regime  at moderate driving amplitudes, and,
further, to developed turbulence at high driving amplitudes (a
Kolmogorov-like cascade \cite{Kolmogorov}  at ${\rm Re} \sim
10^2$), rather than a sharp transition to turbulence at Reynolds
numbers higher than some critical value as occurs in
incompressible fluids \cite{Landau}. In the high frequency domain
a cutoff of the cascade is observed, caused by a change in the
mechanism of energy transfer, from nonlinear wave transformation
to viscous damping. Numerical calculations are in qualitative
agreement with the results of our experiments.

We are grateful to A. A. Levchenko, V. V. Lebedev,  and E. A.
Kuznetsov for valuable discussions.  The investigations were
supported by the Russian Foundation for Basic Research, project
Nos. 05-02-17849  and 06-02-17253, by the Presidium of the Russian
Academy of Sciences under the programs ``Quantum Macrophysics''
and ``Mathematical Methods in Nonlinear Dynamics'', and by the
Engineering and Physical Sciences Research Council (UK).

\end{document}